\documentclass[titlepage,12pt,amstex]{article}
\usepackage{amsfonts}
\usepackage{amssymb}
\usepackage{amsmath}
\usepackage{eurosym}
\usepackage{bbold}
\usepackage{stmaryrd}
\usepackage{times}
\usepackage{booktabs}

\usepackage[utf8]{inputenc}
\usepackage{amssymb}
\usepackage{amsfonts}
\usepackage{amsmath}
\usepackage{mathptmx}
\usepackage{eurosym}
\usepackage{graphicx}
\usepackage{rotating}
\usepackage[justification=centering]{caption}
\usepackage{comment}

\usepackage{ifthen} 
\usepackage{xifthen} 

\usepackage{calc}  
\usepackage{time}  
\usepackage{clock}  

\usepackage{harvard}

\providecommand{\cite}{\citeasnoun}
\renewcommand{\cite}{\citeasnoun}

\usepackage{bibunits}

\usepackage{fancybox}
  
\usepackage{adjustbox} 

\usepackage[justification=centering]{caption} 

\usepackage{section}
\usepackage{theorem}
\usepackage{verbatim}
\providecommand{\bTentative}{}

\usepackage{eurosym} 
\usepackage{fix-cm}  
\usepackage{textcomp} 
\usepackage{pifont}  
\usepackage[T1]{fontenc}
    
\providecommand{\SpecialFonts}{
    \usepackage{calligra} 
	
    \usepackage[varumlaut]{yfonts}  
    }

\usepackage{pdfpages} 

\providecommand{\graphicpackages}{
    \usepackage{epic} 
    \usepackage{eepic}  
    \usepackage{graphpap} 
    \usepackage{pict2e} 
    \usepackage{pictexwd} 
    \usepackage[<pict2e options>]{curve2e}  
    \usepackage{ellipse} 
    \usepackage{pstricks}
    \usepackage{tikz}
    \usepackage{pgfplots}  
  }

\usepackage{gloss}
\usepackage{makeidx} 

\usepackage{enumerate}  
\usepackage{etaremune}  

\usepackage{amssymb}  
\usepackage{amsfonts}  
\usepackage{amsmath}  

\usepackage{array}
\usepackage{bbold} 
\usepackage{dsfont}  
\usepackage{fixmath} 
\usepackage{mathtools}  
\usepackage{mathrsfs} 

\usepackage{endnotes}

\usepackage{afterpage} 
\usepackage[margin=2.5cm]{geometry} 

\usepackage{layout} 
\usepackage{lscape} 

\usepackage{setspace} 

\usepackage{url}  

\usepackage{longtable} 

\usepackage{subfigure}

\providecommand{\paraNumbering}{\theoremstyle{change}}
\SpecialFonts
\excludecomment{ps2pdf}
\includecomment{pdflatex}
\excludecomment{graphicpack}
\excludecomment{Plan}
\excludecomment{Abstract}
\includecomment{AbstractF}
\includecomment{TOCLists}
\includecomment{Calculations}
\includecomment{French}
\includecomment{Alternative}
\includecomment{Old}
\includecomment{Tentative}
\includecomment{Elementary}
\includecomment{Intermediate}
\includecomment{Advanced}
\includecomment{Organization}
\includecomment{Book}
\excludecomment{DP}
\includecomment{DP-W}
\includecomment{DP-CIR}
\includecomment{DP-SSRN}
\includecomment{DP-arXiv}
\includecomment{Lectures}
\includecomment{Article}
\includecomment{Journal}
\includecomment{reversemargin}
\excludecomment{SlidesIn} \includecomment{SlidesOut} \includecomment{Slides}
\includecomment{SlidesLandscapeOut} \includecomment{SlidesLandscape}
\providecommand{\sectitlesize}{}

\newcommand{\mciteA}{\citename}
\newcommand{\mciteY}{\citeyear*}

\newcommand{\mciteAYY}[3][]{\mciteA{#2} (\mciteY{#2}, \mciteY{#3}#1)}

\newcommand{\cbu}{,\xspace}
\newcommand{\dbu}{.\xspace}
\newcommand{\bbibuniteconometrica}{\begin{bibunit}[econometrica] 
    \renewcommand{\cite}{\nocite*}
    \renewcommand\refname{} \renewcommand{\cbu}{} 
    \renewcommand{\dbu}{} \renewcommand{\newline}{}
    \vspace{-3.5\baselineskip}}
\newcommand{\bbibunitagsm}{\begin{bibunit}[agsm] \renewcommand{\cite}{\nocite*}
    \renewcommand\refname{} \renewcommand{\cbu}{} \renewcommand{\dbu}{} 
    \renewcommand{\newline}{}
    \vspace{-3.5\baselineskip}}
\newcommand{\ebibunit}{\putbib[refer1] \end{bibunit}}

\newcommand{\bbibunitunsrt}{\begin{bibunit}[unsrt] \renewcommand{\cite}{\nocite*}
    \renewcommand\refname{} \renewcommand{\cbu}{} \renewcommand{\dbu}{} \renewcommand{\newline}{}
    \vspace{-3.5\baselineskip}}
\newcommand{\ebibunitunsrt}{\putbib[refer1] \end{bibunit}}

\makeindex  
\makeglossary  
\makegloss

\newcommand{\NumericNumberedLists}{
\def\labelenumi{\arabic{enumi}.}
\def\theenumi{\arabic{enumi}}
\def\labelenumii{\arabic{enumii}.}
\def\theenumii{\arabic{enumii}}
\def\p@enumii{\theenumi.}
\def\labelenumiii{\arabic{enumiii}.}
\def\theenumiii{\arabic{enumiii}}
\def\p@enumiii{\theenumi.\theenumii.}
\def\labelenumiv{\arabic{enumiv}.}
\def\theenumiv{\arabic{enumiv}}
\def\p@enumiv{\p@enumiii.\theenumiii}
}

\allowdisplaybreaks

\providecommand{\sectitlesize}{}

\newlength{\totalhormargin}
\newlength{\totalvermargin}

\newcommand{\October}{October }

\def\todayMY{\ifcase\month\or
  January\or February\or March\or April\or May\or June\or
  July\or August\or September\or October\or November\or
  December\fi\ \number\year}

\providecommand{\vartitleadjust}{}
\providecommand{\vartitle}{}

\providecommand{\varAuthors}{Jean-Marie Dufour \thanks{\ \ \DufourAddress} \\
  McGill University}

\providecommand{\vardate}{\today, \texttime}

\providecommand{\DufourAddress}{William Dow Professor of Economics, McGill University,
  Centre interuniversitaire de recherche en analyse des
  organisations (CIRANO), and Centre interuniversitaire de recherche en
  \'{e}conomie quantitative (CIREQ). Mailing address:
  Department of Economics, McGill University, Leacock Building, Room 414,
  855 Sherbrooke Street West, Montr\'{e}al, Qu\'{e}bec H3A 2T7, Canada.
  TEL: (1) 514 398 6071; FAX: (1) 514 398 4800; e-mail: 
  \protect\url=jean-marie.dufour@mcgill.ca=\thinspace. Web page:
  \protect\url{http://www.jeanmariedufour.com} }

\newcommand{\sptha}{\hspace{-0.01em}}
\newcommand{\spthb}{\hspace{-0.01em}}

\newcommand{\sppr}{}
\newcommand{\theoremname}{Theorem}
\newcommand{\acknowledgementname}{Acknowledgement}
\newcommand{\algorithmname}{Algorithm}
\newcommand{\assumptionname}{Assumption}
\newcommand{\axiomname}{Axiom}
\newcommand{\casename}{Case}
\newcommand{\claimname}{Claim}
\newcommand{\conclusionname}{Conclusion}
\newcommand{\conditionname}{Condition}
\newcommand{\conjecturename}{Conjecture}
\newcommand{\corollaryname}{Corollary}
\newcommand{\criterionname}{Criterion}
\newcommand{\definitionname}{Definition}
\newcommand{\examplename}{Example}
\newcommand{\exercisename}{Exercise}
\newcommand{\lemmaname}{Lemma}
\newcommand{\notationname}{Notation}
\newcommand{\problemname}{Problem}
\newcommand{\proofname}{\sppr Proof}

\newcommand{\propertyname}{\sppr Property}
\newcommand{\propositionname}{Proposition}
\newcommand{\reflistname}{References}
\newcommand{\remarkname}{Remark}
\newcommand{\resultname}{Result}
\newcommand{\solutionname}{Solution}

\newcommand{\summaryname}{Summary}

\providecommand{\proofof}{Proof of }
\renewcommand{\proofof}{Proof of }

\newenvironment{proof}[1][\sptha \proofname]{\par
  \normalfont
  \trivlist
  \item[\hskip\labelsep\scshape
    #1{.}]\ignorespaces}
    {\qed\endtrivlist\vspace{\baselineskip}}

\newenvironment{proofnoname}[1][\sptha \proofname]{\par
  \noindent
  \normalfont}%
  {\qed\vspace{\baselineskip}}

\newenvironment{proofflex}[1][\spthb \proofname]{\par
  \normalfont
  \trivlist
  \item[\hskip\labelsep\scshape
    #1{\ }]\ignorespaces}
    {\qed\endtrivlist\vspace{\baselineskip}}

\newenvironment{proofflexb}[1][\spthb \proofname]{\par
  \normalfont
  \trivlist
  \item[\hskip\labelsep\scshape
    #1{\ }]\ignorespaces}
    {\qed\endtrivlist\vspace{\baselineskip}}

\newenvironment{proofflexc}[1][\spthb \proofname]{\par
  \noindent
  \normalfont}
  {\qed\vspace{\baselineskip}}

\newenvironment{proofflexwc}[1][\spthb \proofname]{\par
  \normalfont
  \trivlist
  \item[\hskip\labelsep\scshape
    #1{\ }]\ignorespaces}

\newenvironment{proofflexd}[1][\spthb \proofname]{\par
  \noindent
  \normalfont}
  {\vspace{-0.5\baselineskip}}

\newenvironment{sec2subsec}{\renewcommand{\section}{\subsection}}{}
\newenvironment{reflist}{\quad \newline \noindent {\Large \bf \reflistname}
  \newline \quad \newline
  \begin{list}{}{\itemindent -.15in \leftmargin .15in \parsep 0in
                 \itemsep 0in} \item \vspace{-.35in} }{\end{list}}

\newenvironment{npar}{\noindent \bf{\thesection.\thenpar}}{}
\newenvironment{subnpar}{\noindent \bf{\thesubsection.\thesubnpar}}{}

\newcounter{paran}

\newcounter{npar}[section]
\newcounter{nparr}[section]

\newcounter{subnpar}[subsection]
\newcounter{subnparr}[subsection]

\DeclareRobustCommand{\qed}{%
  \ifmmode 
  \else \leavevmode\unskip\penalty9999 \hbox{}\nobreak\hfill
  \fi
  \quad\hbox{\qedsymbol}}
\newcommand{\openbox}{\leavevmode
  \hbox to.77778em{%
  \hfil\vrule
  \vbox to.675em{\hrule width.6em\vfil\hrule}%
  \vrule\hfil}}
\DeclareRobustCommand{\qeddirect}{%
  \ifmmode 
  \else \leavevmode\unskip\penalty9999 \hbox{}\nobreak\hfill
  \fi
  \quad\hbox{\qedsymbol}}
\providecommand{\qedsymbol}{\openbox} 

\newcommand{\bproofin}{\begin{proof}}
\newcommand{\eproofin}{\end{proof}}
\newcommand{\bproofend}{\begin{proofnoname}}
\newcommand{\eproofend}{\end{proofnoname}}
\newcommand{\bproofth}{\begin{proofth}}
\newcommand{\eproofth}{\end{proofth}}

\providecommand{\proofof}{Proof of }
\renewcommand{\proofof}{Proof of }

\newcommand{\thsection}{\thesection}

\newcommand{\thsectioneq}{\thesection}

\newcommand{\thsec}{\thsection}
\newcommand{\thseceq}{\thsectioneq}

\renewcommand{\theequation}{{\rm \thseceq.\arabic{equation}}}

\makeatletter
   \long
\def\@makecaption#1#2{\vskip 0\p@
   \setbox\@tempboxa\hbox{#1 #2}}
\makeatother

\makeatletter

\providecommand{\l@theorem}{\@dottedtocline{1}{0em}{5em}}
\renewcommand{\l@theorem}{\@dottedtocline{1}{0em}{5em}}

\providecommand{\listttheoremnameb}{\sectitlesize List of Definitions, Assumptions, Propositions and Theorems}

\newcommand\listoftheorems{
 \section*{\listttheoremnameb
           \@mkboth{\MakeUppercase\listttheoremnameb}
           {\MakeUppercase\listttheoremnameb}}
           \@starttoc{lth}
           }
\newcommand{\listoftheoremscont}[1]{
 \providecommand{\listttheoremnameb}{#1}
 \renewcommand{\listttheoremnameb}{#1}
 \section*{\listttheoremnameb
           \@mkboth{\MakeUppercase\listttheoremnameb}
           {\MakeUppercase\listttheoremnameb}}
           \@starttoc{lth}
                      \addcontentsline{toc}{section}{\listttheoremnameb}
           }

\makeatother
\providecommand{\contentshift}{\hspace{-3.5em}}
\renewcommand{\contentshift}{\hspace{-3.5em}}
\providecommand{\contentshiftS}{\hspace{0em}}
\renewcommand{\contentshiftS}{\hspace{0em}}

\newcommand{\sppra}{\hspace{-0.2em}}
\newcommand{\indentprc}{}

\newcommand{\captionproofflex}[2]{}

\newcommand{\captionproofflexc}[2]{
 \sppra\indentprc\textsc{Proof of #1 \protect\ref{#2} }\quad
 \addcontentsline{lth}{theorem}{\protect\numberline{}
  {\contentshift \proofof #1 \protect\ref{#2} }}}

\newcommand{\bAssumptionA}{\begin{assumption}}
\newcommand{\eAssumptionA}{\end{assumption}}

\providecommand{\paraNumbering}{}
\renewcommand{\paraNumbering}{}
\paraNumbering

\newtheorem{theorem}{\sptha \theoremname}[section]
\newtheorem{theoremseq}{\sptha \theoremname}
{\theorembodyfont{\normalfont}%
 \newtheorem{theoremSeqRom}{\sptha \theoremname}}
\newtheorem{acknowledgement}{\sptha \acknowledgementname}[section]
\newtheorem{algorithm}{\sptha \algorithmname}[section]
\newtheorem{assumption}{\sptha \assumptionname}[section]
\newtheorem{assumptionSec}{\sptha \assumptionname}[section]
\newtheorem{assumptionV}{\sptha \assumptionname}[section]
\newtheorem{assumptionLetter}{\sptha \assumptionname}

\newtheorem{axiom}{\sptha \axiomname}[section]
\newtheorem{case}{\sptha \casename}[section]
\newtheorem{claim}{\sptha \claimname}[section]
\newtheorem{conclusion}{\sptha \conclusionname}[section]
\newtheorem{condition}{\sptha \conditionname}[section]
\newtheorem{conjecture}{\sptha \conjecturename}[section]
\newtheorem{corollary}[theorem]{\sptha \corollaryname}
\newtheorem{criterion}{\sptha \criterionname}[section]
\newtheorem{definition}{\sptha \definitionname}[section]
\newtheorem{definitionSec}{\sptha \definitionname}[section]
{\theorembodyfont{\normalfont}%
    \newtheorem{example}{\sptha \examplename}[section]}
{\theorembodyfont{\normalfont}
    \newtheorem{exampleSec}{\sptha \examplename}[section]}
\newtheorem{exercise}{\sptha \exercisename}[section]
\newtheorem{lemma}[theorem]{\sptha \lemmaname}
\newtheorem{lemmaSec}{\sptha \lemmaname}[section]
\newtheorem{notation}{\sptha \notationname}[section]
\newtheorem{problem}{\sptha \problemname}[section]
{\theorembodyfont{\normalfont} 
    \newtheorem{proofth}{\sptha \proofname}[section]}
\newtheorem{property}[theorem]{\sptha \propertyname}
\newtheorem{proposition}[theorem]{\sptha \propositionname}
\newtheorem{propositionSec}{\sptha \propositionname}[section]
{\theorembodyfont{\normalfont} 
    \newtheorem{remark}{\sptha \remarkname}[section]}
    {\theorembodyfont{\normalfont}%
\newtheorem{result}{\sptha \resultname}[section]}
\newtheorem{solution}{\sptha \solutionname}[section]
\newenvironment{statement}{}{}
\newtheorem{summary}{\sptha \summaryname}[section]

\providecommand{\resetcountersSection}{\renewcommand{\thsec}{\thsection} 
  \renewcommand{\thseceq}{\thsectioneq} 
  \setcounter{theorem}{0} \setcounter{definition}{0} \setcounter{equation}{0}}

\newcommand{\captionproposition}[2]{\textsc{ #2}.
\addcontentsline{lth}{theorem}{{\bf #1 \contentshiftS \protect\numberline{\theproposition}}
 {\contentshift : \contentshiftS #2}}}

\newcommand{\captionproofemptynocontent}[2]{}

\newcommand{\captionproofin}[2]{}

\providecommand{\vartitleadjust}{}
\renewcommand{\vartitleadjust}{}

\providecommand{\vartitle}{\vartitleadjust Gini representations of moments, decoupling and generalized Lagrange 
                identities}
\providecommand{\vartitle}{\vartitleadjust Pairwise representations of moments, $U$-statistics 
                and generalized Lagrange identities}
\providecommand{\vartitle}{\vartitleadjust Pairwise difference representations of moments: Gini, Cauchy-Schwarz 
              and generalized Lagrange identities}
\renewcommand{\vartitle}{\vartitleadjust Pairwise difference representations of moments: \\ 
                         Gini and generalized Lagrange identities}
\renewcommand{\vartitle}{\vartitleadjust Pairwise Difference Representations of Moments: \\ 
	Gini and Generalized Lagrange identities}
\renewcommand{\vartitle}{\vartitleadjust Gini representations of skewness, kurtosis and \\ 
	higher-order moments}

\providecommand{\varAuthors}{Jean-Marie Dufour \thanks{\ \ \DufourAddress} \\
                McGill University \and Abderrahim Taamouti \thanks{\ \ \TaamoutiAAddress}\\ Universidad Carlos III de Madrid}
\renewcommand{\varAuthors}{Jean-Marie Dufour \thanks{\ \ \DufourAddress} \\ McGill University 
              \and Abderrahim Taamouti \thanks{\ \ \TaamoutiAAddress}\\ University of Liverpool}
\renewcommand{\varAuthors}{Jean-Marie Dufour \thanks{\ \ \DufourAddress} \\ McGill University 
              \and Abderrahim Taamouti \thanks{\ \ \TaamoutiAAddress}\\ University of Liverpool \\
			  \and Meilin Tong \thanks{\ \ \TongAddress} \\  McGill University}
\providecommand{\vardate}{\today, \texttime}
\providecommand{\vardate}{\firstversion{\February 2012} \\ \revised{\August 2012, \August 2012, \January 2024} 
     \\ \thisversion{\January 2024} \\ \compiled{\today, \texttime}}
\renewcommand{\vardate}{\firstversion{\February 2012} \\ \revised{\August 2012, \January 2024} 
     \\ \thisversion{\January 2024} \\ \compiled{\today, \texttime}}
\renewcommand{\vardate}{\April 2024 \\ \compiled{\today, \texttime}}

\providecommand{\listttheoremnameb}{List of Definitions, Assumptions, Propositions 
     and Theorems}
\renewcommand{\listttheoremnameb}{List of Definitions, Assumptions, Propositions 
     and Theorems}
\renewcommand{\thefootnote}{\alph{footnote}}

\def \beginenumerateT {\begin{enumerate}}
\def \endenumerateT {\end{enumerate}}

\def \beginenumerateTh[#1] {\begin{enumerate}[#1]}
\def \endenumerateTh {\end{enumerate}}

\begin{French}
\end{French}
\begin{graphicpack}
\end{graphicpack}
\begin{DP}
\end{DP}
\begin{DP-W}
\end{DP-W}
\begin{DP-CIR}
\end{DP-CIR}
\begin{DP-SSRN}
\end{DP-SSRN}
\begin{DP-arXiv}
\end{DP-arXiv}
\begin{Lectures} 
\end{Lectures}
\begin{Organization}
\end{Organization}
\begin{reversemargin}
\end{reversemargin}
\begin{SlidesIn}

\providecommand{\sectitlesize}{\huge}
\renewcommand{\sectitlesize}{\huge}

\providecommand{\vartitleadjust}{\Huge}
\renewcommand{\vartitleadjust}{\Huge}

\end{SlidesIn}
\begin{SlidesOut}
\end{SlidesOut}
\begin{Slides}
\end{Slides}
\begin{SlidesLandscapeOut}
\end{SlidesLandscapeOut}
\begin{Article} 
\end{Article}
\begin{Journal}
\end{Journal}
\input{tcilatex}

\renewcommand{\vardate}{\October 2026}
\input{tcilatex}
\begin{document}

\title{%
\vartitle 
}
\author{%
\varAuthors%
}
\date{%
\vardate%
}
\maketitle

\begin{Slides}
\end{Slides}

\begin{Tentative}
\end{Tentative}

\newpage

\pagestyle{plain}\pagenumbering{roman}\setcounter{page}{1}

\begin{center}
\textbf{ABSTRACT}

\quad
\end{center}

\noindent We provide pairwise-difference (Gini-type) representations of
higher-order central moments for both general random variables and empirical
moments. Such representations do not require a measure of location. For
third and fourth moments, this yields pairwise-difference representations of
skewness and kurtosis coefficients. We show that all central moments possess
such representations, so no reference to the mean is needed for moments of
any order. This is done by considering i.i.d. \emph{replications} of the
random variables considered, by observing that central moments can be
interpreted as covariances between a random variable and powers of the same
variable, and by giving recursions which link the pairwise-difference
representation of any moment to lower order ones. 

\quad

\vspace{0.25pt}

\quad

\begin{Tentative}
\end{Tentative}

\newpage

\pagenumbering{arabic} \setcounter{section}{0} \setcounter{page}{1} 
\renewcommand{\thefootnote}{\arabic{footnote}}%

\section{\sectitlesize Introduction \label{Sec: Introduction}}

\resetcountersSection

The sample variance is certainly the most widely used measure of dispersion
among observations in statistics. It is typically defined as the average of
the squared deviations of the observations $x_{1},\ldots ,\,x_{n}$ from
their sample mean:%
\begin{equation}
s_{X}^{2}(n):=\frac{1}{n-1}\overset{n}{\underset{i=1}{\sum }}(x_{i}-\bar{x}%
)^{2}  \label{eq: Sample variance}
\end{equation}%
where $\bar{x}=\overset{n}{\underset{i=1}{\sum }}x_{i}/n$. At first sight,
the variance depends crucially on \textquotedblleft
centering\textquotedblright\ the observations with respect to their mean. A
century ago, however, \cite{Gini(1912)} noticed that $s_{X}^{2}(n)$ can be
rewritten as the average of pairwise differences between all the
observations: 
\begin{equation}
s_{X}^{2}(n)=\frac{1}{2}\frac{1}{n(n-1)}\overset{n}{\underset{i=1}{\sum }}%
\overset{n}{\underset{j=1}{\sum }}(x_{i}-x_{j})^{2}=\frac{1}{n(n-1)}\underset%
{i=1}{\overset{n}{\sum }}\,\underset{j>i}{\sum }(x_{i}-x_{j})^{2}\,;
\label{eq: Pairwise sample variance}
\end{equation}%
see \cite{Heffernan(1988)}. The Gini representation underscores that $%
s_{X}^{2}(n)$ does not depend on using the sample mean as location
parameter, and thus may be viewed as a measure of \textquotedblleft
intrinsic variability\textquotedblright\ between observations. In
particular, it provides a way of measuring dispersion when there is no
natural location parameter, as with categorical data; see 
\mciteAYY{Light-Margolin(1971)}{Light-Margolin(1974)}%
. Further, by focusing on pairwise differences $x_{i}-x_{j}$, it may be
easier to determine which observations have the greatest influence on $%
s_{X}^{2}(n)$, because deviations from the mean $x_{i}-\bar{x}$ depend on
all the observations (through $\bar{x})$. A comprehensive review of
statistical methods based on the Gini approach is presented by \cite%
{Yitzhaki-Schechtman(2013)}.

If we have bivariate observations $z_{i}:=(x_{i},$\thinspace $y_{i})^{\prime
}$, $i=1,\ldots ,\,n,$ the sample covariance 
\begin{equation}
s_{XY}(n):=\frac{1}{n-1}\overset{n}{\underset{i=1}{\sum }}(x_{i}-\bar{x}%
)(y_{i}-\bar{y})  \label{eq: Sample covariance}
\end{equation}%
is in turn the most widely used \textquotedblleft measure of
association\textquotedblright\ between $x$ and $y.$ For the covariance, a
pairwise representation is provided by the formula%
\begin{equation}
s_{XY}(n)=\frac{1}{2}\frac{1}{n(n-1)}\overset{n}{\underset{i=1}{\sum }}%
\overset{n}{\underset{j=1}{\sum }}(x_{i}-x_{j})(y_{i}-y_{j})=\frac{1}{n(n-1)}%
\overset{n}{\underset{i=1}{\sum }}\underset{j>i}{\sum }%
(x_{i}-x_{j})(y_{i}-y_{j})\,;  \label{eq: Pairwise sample covariance}
\end{equation}%
see \cite{Hayes(2011)}. Clearly, (\ref{eq: Pairwise sample variance}) is a
special case of (\ref{eq: Pairwise sample covariance}) obtained by taking $%
x_{i}=y_{i},\;i=1,\ldots ,\,n.$ Formula (\ref{eq: Pairwise sample covariance}%
) shows that the sample covariance measures the tendency of $x$ and $y$ to
move in the same direction, without reference to sample means. Similar
expressions for the variances and covariances of general random variables
have also been used in the literature on $U$-statistics; see \cite[Chapter 1]%
{LeeAJ(1990)}. Since centering is not needed, formula (\ref{eq: Pairwise
sample covariance}) provides a basis for developing alternative measures of
dependence, such as Kendall's and Spearman's measures of dependence [see 
\cite{Lehmann(1966)}]. Again, pairwise differences $x_{i}-x_{j}$ and $%
y_{i}-y_{j}$ may be easier to interpret than $x_{i}-\bar{x}$ and $y_{i}-\bar{%
y}$, because $\bar{x}$ and $\bar{y}$ depend on all the observations.

Since the representations in (\ref{eq: Pairwise sample variance}) and (\ref%
{eq: Pairwise sample covariance}) only depend on \emph{differences} between
observations, we will call such representations (and similar ones) $D$\emph{%
-representations }(or \emph{Gini-type representations}). In this paper, we
provide $D$\emph{-}representations of higher-order central moments for both
general random variables and empirical moments. For third and fourth
moments, this yields new representations of skewness and kurtosis
coefficients, where reference to a location parameter is eliminated. We show
that all central moments possess $D$\emph{-}representations, so again no
reference to the mean is needed for moments of any order. This is done by
considering i.i.d. \emph{replications} (or realizations) of the random
variables considered, by observing that central moments can be interpreted
as covariances between a random variable and powers of the same variable (a 
\emph{nonlinear} relation), and by giving \emph{recursions} which link the $%
D $-representation of any moment to lower order ones. Numerical summation
identities similar to (\ref{eq: Pairwise sample variance})\thinspace
-\thinspace (\ref{eq: Pairwise sample covariance}) are also deduced.

The paper is organized as follows. In Section \ref{Sec: Gini representation
of covariances}, we discuss $D$-representations for covariances and
regression coefficients, and we show how corresponding numerical identities
can be derived from these. In Section \ref{Sec: Skewness and kurtosis}, we
provide $D$-representations for third and fourth moments, along with
corresponding expressions for skewness and kurtosis coefficients. In Section %
\ref{Sec: Higher-order moments}, we show that all moments possess such
representations, and we give recursive formulae for computing a $D$%
-representation for any moment. We conclude in Section \ref{Sec: Conclusion}%
. Proofs are given in Appendix.

\section{$D$-representation of Covariances \label{Sec: Gini representation
of covariances}}

\resetcountersSection

In order to generalize the expressions (\ref{eq: Pairwise sample variance}%
)\thinspace -\thinspace (\ref{eq: Pairwise sample covariance}) to
higher-order moments, it will be useful to discuss the pairwise difference
representation of the covariance between general random variables. Let $X$
and $Y$ be random variables with finite second moments, and means $\mu _{X}=%
\mathbb{E}\{X\}$, $\mu _{Y}=\mathbb{E}\{Y\}.$ We denote by $\mathsf{C}%
[X,\,Y]:=\mathbb{E}\{(X-\mu _{X})(Y-\mu _{Y})\}$ the covariance between $X$
and $Y$. Consider two independent and identically distributed replications
of the random vector $(X,\,Y)^{\prime }$: $\;(X_{1},\,Y_{1})^{\prime }$ and $%
(X_{2},\,Y_{2})^{\prime }$. On observing 
\begin{eqnarray}
\mathbb{E}\{(X_{1}-X_{2})(Y_{1}-Y_{2})\} &=&\mathbb{E}\{X_{1}Y_{1}\}+\mathbb{%
E}\{X_{2}Y_{2}\}-\mathbb{E}\{X_{1}Y_{2}\}-\mathbb{E}\{X_{2}Y_{1}\}  \notag \\
&=&\mathbb{E}\{X_{1}Y_{1}\}+\mathbb{E}\{X_{2}Y_{2}\}-\mathbb{E}\{X_{1}\}%
\mathbb{E}\{Y_{2}\}-\mathbb{E}\{X_{2}\}\mathbb{E}\{Y_{1}\}  \notag \\
&=&2(\mathbb{E}\{XY\}-\mathbb{E}\{X\}\mathbb{E}\{Y\})=2\mathsf{C}[X,\,Y]\,,
\label{eq: E[(X_1-X_2)(Y_1-Y_2)]}
\end{eqnarray}%
it follows that%
\begin{equation}
\mathsf{C}[X,\,Y]=\frac{1}{2}\mathbb{E}\{(X_{1}-X_{2})(Y_{1}-Y_{2})\}\,.
\label{eq: Cov}
\end{equation}%
It is also easy to see that%
\begin{eqnarray}
\mathsf{C}[X,\,Y] &=&\mathbb{E}\{X_{1}(Y_{1}-Y_{2})\}=-\mathbb{E}%
\{X_{2}(Y_{1}-Y_{2})\}  \notag \\
&=&\mathbb{E}\{(X_{1}-X_{2})Y_{1}\}=-\mathbb{E}\{(X_{1}-X_{2})Y_{2}\}\,.
\label{eq: Cov One diff}
\end{eqnarray}%
In other words, only one of the variables in $(X,$\thinspace $Y)^{\prime }$
need be replicated to avoid centering with respect to the means ($\mu _{X}$
and $\mu _{Y}$). (\ref{eq: E[(X_1-X_2)(Y_1-Y_2)]})\thinspace -\thinspace (%
\ref{eq: Cov One diff}) also hold under the weaker assumption (without
identical distributions) that $(X_{1},\,Y_{1})^{\prime }$ and $%
(X_{2},\,Y_{2})^{\prime }$ have the same first and second moments as $%
(X,\,Y) $ with 
\begin{equation}
\mathbb{E}\{X_{1}Y_{2}\}=\mathbb{E}\{X_{2}Y_{1}\}=\mathbb{E}\{X\}\mathbb{E}%
\{Y\}\,.
\end{equation}

For $X_{1}=Y_{1}$ and $X_{2}=Y_{2},$ we obtain the following formulae for
the variance of $X:$%
\begin{eqnarray}
\sigma _{X}^{2} &=&\mathrm{Var}(X)=\frac{1}{2}\mathbb{E}\{(X_{1}-X_{2})^{2}\}
\notag \\
&=&\mathbb{E}\{X_{1}(X_{1}-X_{2})\}=-\mathbb{E}\{X_{2}(X_{1}-X_{2})\}\,.
\label{eq: Var}
\end{eqnarray}%
If $X_{3}$ is a third replication of $X$ [so that $X_{1}$, $X_{2}$, $X_{3}$
are i.i.d.], it is also easy to check that 
\begin{equation}
\sigma _{X}^{2}=\mathbb{E}\{(X_{1}-X_{3})(X_{1}-X_{2})\}\,.
\label{eq: Var 3 rep}
\end{equation}%
The uncentered second moment of $X$ can be written as%
\begin{eqnarray}
\mathbb{E}\{X^{2}\} &=&\frac{1}{2}\mathbb{E}\{(X_{1}-X_{2})^{2}\}+\mu
_{X}^{2}  \notag \\
&=&\mathbb{E}\{X_{1}(X_{1}-X_{2})\}+\mu _{X}^{2}=-\mathbb{E}%
\{X_{2}(X_{1}-X_{2})\}+\mu _{X}^{2}
\end{eqnarray}%
and the linear regression coefficient of $Y$ on $X$ is%
\begin{eqnarray}
\beta _{Y\cdot X} &=&\frac{\sigma _{XY}}{\sigma _{X}^{2}}=\frac{\mathbb{E}%
\{(X_{1}-X_{2})(Y_{1}-Y_{2})\}}{\mathbb{E}\{(X_{1}-X_{2})^{2}\}}  \notag \\
&=&\frac{\mathbb{E}\{(X_{1}-X_{2})Y_{1}\}}{\mathbb{E}\{(X_{1}-X_{2})X_{1}\}}=%
\frac{\mathbb{E}\{X_{1}(Y_{1}-Y_{2})\}}{\mathbb{E}\{X_{1}(X_{1}-X_{2})\}}\,.
\end{eqnarray}

In view of the above discussion on the covariance and the variance, we will
define more formally what we mean by the $D$-representation of a parameter.

\begin{definition}
A parameter $\gamma $ has a $D$-representation for the variables $%
X_{1},\ldots ,\,X_{n}$ if there is a function $g(x_{1},\ldots
,\,x_{n})=h(x_{2}-x_{1},\ldots ,\,x_{n}-x_{n-1})$ such that 
\begin{equation}
\gamma =\mathbb{E}\{h(X_{2}-X_{1},\ldots ,\,X_{n}-X_{n-1})\}\,.
\end{equation}
\end{definition}

In other words, $\gamma $ has a $D$-representation for the variables $%
X_{1},\ldots ,\,X_{n}$ if it can be written as the expected value of a
function $g(X_{1},\ldots ,\,X_{n})$ which depends on $X_{1},\ldots ,\,X_{n}$
only through the differences between the variables $X_{1},\ldots ,\,X_{n}$.
Equivalently, this means that 
\begin{equation}
\gamma =\mathbb{E}\{g(X_{1},\ldots ,\,X_{n})\}
\end{equation}%
where $g(x_{1},\ldots ,\,x_{n})$ is invariant to uniform location changes, 
\emph{i.e.}%
\begin{equation}
g(x_{1}+c,\ldots ,\,x_{n}+c)=g(x_{1},\ldots ,\,x_{n})\quad \text{for any }c.
\end{equation}

A potentially useful feature of the representation (\ref{eq: Cov}) follows
from the fact that the variables $(X_{1}-X_{2})$ and $(Y_{1}-Y_{2})$ have
marginal distributions symmetric around zero. Consequently, the odd moments
of $(X_{1}-X_{2})$ and $(Y_{1}-Y_{2})$ are all zero (when they exist). 
Further, the distribution of the vector $(X_{1}-X_{2},\,Y_{1}-Y_{2})^{%
\prime }$ is jointly symmetric around zero, \textit{i.e.}
\begin{equation}
(X_{1}-X_{2},\,Y_{1}-Y_{2})^{\prime }\sim
-(X_{1}-X_{2},\,Y_{1}-Y_{2})^{\prime }.
\end{equation}%
This type of symmetry can be useful for proving unbiasedness results in
various setups; see \mciteAYY{Dufour(1984)}{Dufour(1985)}. This property is
shared by all $D$-representations when the paired differences are based on
i.i.d. observations.

It is of interest to note that the numerical identity (\ref{eq: Pairwise
	sample covariance}) can be derived from the general moment formula (\ref{eq:
	Cov}). Let $I$ be a random integer such that\ $\mathbb{P}[I=i]=1/n\,,%
\;i=1,...,n$, and consider the random vector\ $(X^{\ast },\,Y^{\ast
})=(x_{I},\,y_{I})$. The values $x_{1},\ldots ,\,x_{n}$ and $y_{1},\ldots
,\,y_{n}$ need not be distinct. It is then easy to see that:%
\begin{equation}
	\mathbb{E}\{X^{\ast }Y^{\ast }\}=\frac{1}{n}\sum_{i=1}^{n}x_{i}y_{i}\,,\quad 
	\mathbb{E}\{X^{\ast }\}=\frac{1}{n}\sum_{i=1}^{n}x_{i}=\bar{x}\,,\quad 
	\mathbb{E}\{Y^{\ast }\}=\frac{1}{n}\sum_{i=1}^{n}y_{i}=\bar{y}\,,
	\label{eq: E(X*Y*)}
\end{equation}%
\begin{equation}
	\sigma _{X^{\ast }Y^{\ast }}=\mathbb{E}\{X^{\ast }Y^{\ast }\}-\mathbb{E}%
	\{X^{\ast }\}\mathbb{E}\{Y^{\ast }\}=\frac{1}{n^{2}}\sum_{i=1}^{n}%
	\sum_{j=1}^{n}x_{i}y_{j}-\bar{x}\bar{y}\,.
\end{equation}%
Consider now two i.i.d. replications of $(X^{\ast },\,Y^{\ast })$:\ \ $%
(X_{1}^{\ast },\,Y_{1}^{\ast })$ and $(X_{2}^{\ast },\,Y_{2}^{\ast })$. By
applying (\ref{eq: Cov}) and (\ref{eq: E(X*Y*)}) to $(X^{\ast },$\thinspace $%
Y^{\ast })$, the covariance between $X^{\ast }$ and $Y^{\ast }$ is given by:%
\begin{equation}
	\sigma _{X^{\ast }Y^{\ast }}=\frac{1}{2}\mathbb{E}\{(X_{1}^{\ast
	}-X_{2}^{\ast })(Y_{1}^{\ast }-Y_{2}^{\ast })\}=\frac{1}{2}\frac{1}{n^{2}}%
	\sum_{i=1}^{n}\sum_{j=1}^{n}(x_{i}-x_{j})(y_{i}-y_{j})\,.
\end{equation}%
Since the above sum contains a zero whenever $i=j$, it is natural to divide
by $(n^{2}-n)\;$rather than $n^{2}$. This leads to: 
\begin{equation}
s_{X^{\ast }Y^{\ast }}:=(\frac{n}{n-1})\sigma _{X^{\ast }Y^{\ast }}=\frac{1}{%
2}\frac{1}{n(n-1)}\sum_{i=1}^{n}%
\sum_{j=1}^{n}(x_{i}-x_{j})(y_{i}-y_{j})=s_{XY}(n)\,.
\end{equation}%
In the special case where $X^{\ast }=Y^{\ast },$ we have%
\begin{equation}
\sigma _{X^{\ast }}^{2}=\frac{1}{2}\mathbb{E}\{(X_{1}^{\ast }-X_{2}^{\ast
})^{2}\}=\frac{1}{2}\frac{1}{n^{2}}\sum_{i=1}^{n}%
\sum_{j=1}^{n}(x_{i}-x_{j})^{2}
\end{equation}%
hence, on setting $s_{X^{\ast }}^{2}:=(\frac{n}{n-1})\sigma _{X^{\ast }}^{2}$%
, the Gini formula (\ref{eq: Pairwise sample variance}) for the variance is:%
\begin{equation}
s_{X^{\ast }}^{2}=\frac{1}{2}\frac{1}{n(n-1)}\sum_{i=1}^{n}%
\sum_{j=1}^{n}(x_{i}-x_{j})^{2}=s_{X}^{2}(n)\,.
\end{equation}

\section{Skewness and Kurtosis \label{Sec: Skewness and kurtosis}}

\resetcountersSection

In this section, we express skewness and kurtosis coefficients in terms of
pairwise differences. Such representations can be written as functions of
differences of the form $X_{i}-X_{j}$ or differences between powers of
variables [$X_{i}^{2}-X_{j}^{2}$]. We first underscore that the third and
fourth central moments of a random variable $X$ can be interpreted as
covariances, which in turn can be represented in terms of pairwise
differences between replicated random variables. In each case, we give
several expressions which may have independent interest.

\begin{proposition}
\label{Th: Covariance representation of third and fourth central moments} 
\captionproposition{Proposition}{Covariance representation of third and
fourth central moments} Let $X$ be a random variable with finite mean $\mu
_{X}$ and variance $\sigma _{X}^{2}$. Let $X_{1}$, $X_{2}$ two \emph{i.i.d.}
replications of $X$. If\ $X$ has finite third moment, then%
\begin{eqnarray}
\mathbb{E}\{(X-\mu _{X})^{3}\} &=&\mathsf{C}[X,\,(X-\mu _{X})^{2}]=\mathsf{C}%
[X,\,X^{2}]-2\mu _{X}\sigma _{X}^{2}  \notag \\
&=&\mathbb{E}\{(X_{1}-X_{2})(X_{1}-\mu _{X})^{2}\}=\mathbb{E}%
\{(X_{1}-X_{2})X_{1}^{2}\}-2\mu _{X}\sigma _{X}^{2}  \notag \\
&=&\frac{1}{2}\mathbb{E}\{(X_{1}-X_{2})[(X_{1}-\mu _{X})^{2}-(X_{2}-\mu
_{X})^{2}]\}  \notag \\
&=&\frac{1}{2}\mathbb{E}\{(X_{1}-X_{2})(X_{1}^{2}-X_{2}^{2})\}-\mu _{X}%
\mathbb{E}\{(X_{1}-X_{2})^{2}\}\,.  \label{eq: Cov Third moment}
\end{eqnarray}%
If $X$ has finite fourth moment, then%
\begin{eqnarray}
\mathbb{E}\{(X-\mu _{X})^{4}\} &=&\mathsf{C}[X,\,(X-\mu _{X})^{3}]=\mathsf{C}%
[X,\,X^{3}]-3\mu _{X}\mathsf{C}[X,\,X^{2}]+3\mu _{X}^{2}\sigma _{X}^{2} 
\notag \\
&=&\mathbb{E}\{(X_{1}-X_{2})(X_{1}-\mu _{X})^{3}\}=\frac{1}{2}\mathbb{E}\{(X_{1}-X_{2})[(X_{1}-\mu _{X})^{3}-(X_{2}-\mu
_{X})^{3}]\}  \notag \\
&=&\frac{1}{2}\mathbb{E}\{(X_{1}-X_{2})(X_{1}^{3}-X_{2}^{3})\}-3\mu _{X}%
\mathbb{E}\{(X-\mu _{X})^{3}\}-3\mu _{X}^{2}\sigma _{X}^{2}\,.
\label{eq: Cov Fourth moment}
\end{eqnarray}
\end{proposition}

To interpret Proposition \ref{Th: Covariance representation of third and
fourth central moments}, consider first the case where the mean is zero ($%
\mu _{X}=0$), so the third moment does not depend on the mean or the
variance. When the distribution is skewed to the right [$\mathbb{E}\{(X-\mu
_{X})^{3}\}>0$\}, equation (\ref{eq: Cov Third moment}) shows that an
increase in the value of $X$ ($X_{1}-X_{2}>0$) is associated with an
increase in the absolute value of $X$ [$X_{1}^{2}-X_{2}^{2}>0$], while a
shift to the left would go with a decrease in the absolute value of $X$.
More generally, $\mathbb{E}\{(X-\mu _{X})^{3}\}$ can be written as%
\begin{equation}
\mathbb{E}\{(X-\mu _{X})^{3}\}=\frac{1}{2}\mathbb{E}\{[(X_{1}-\mu
_{X})-(X_{2}-\mu _{X})][(X_{1}-\mu _{X})^{2}-(X_{2}-\mu _{X})^{2}]\}\,
\end{equation}%
so that $\mathbb{E}\{(X-\mu _{X})^{3}\}$ measures the association between a
change in the deviation from the mean and the corresponding change in the
absolute values of the deviations. A positive (negative) association entails
positive (negative) skewness, which is clearly exhibited by pairwise
differences.

The formulae given by Proposition \ref{Th: Covariance representation of
third and fourth central moments} depend on unknown parameters ($\mu _{X}$
and $\sigma _{X}^{2}$). It is possible to avoid this by using a larger
number of replications. For the third moment, only one additional
replication of $X$ is needed. If $X_{1}$, $X_{2}$, $X_{3}$ are three i.i.d.
replications of $X$, we see [from (\ref{eq: Cov Third moment})] that%
\begin{eqnarray}
\mathbb{E}\{(X-\mu _{X})^{3}\} &=&\frac{1}{2}\mathbb{E}%
\{(X_{1}-X_{2})(X_{1}^{2}-X_{2}^{2})\}-\mathbb{E}\{X_{3}\}\mathbb{E}%
\{(X_{1}-X_{2})^{2}\}  \notag \\
&=&\frac{1}{2}\mathbb{E}%
\{(X_{1}-X_{2})[(X_{1}^{2}-X_{2}^{2})-2X_{3}(X_{1}-X_{2})]\}\,.
\label{eq: Cov Third moment 3 rep}
\end{eqnarray}%
For the fourth moment, this can be done with four replications. If $X_{1}$, $%
X_{2}$, $X_{3}$, $X_{4}$ are four i.i.d. replications of $X$, we have: 
\begin{align*}
\mathbb{E}\{(X-\mu _{X})^{4}\}=& \frac{1}{2}\mathbb{E}%
\{(X_{1}-X_{2})(X_{1}^{3}-X_{2}^{3})\}-\frac{3}{2}\mathbb{E}\{X_{3}\}\mathbb{%
E}\{(X_{1}-X_{2})(X_{1}^{2}-X_{2}^{2})\} \\
& +\frac{3}{2}\mathbb{E}\{X_{3}\}\mathbb{E}\{X_{4}\}\mathbb{E}%
\{(X_{1}-X_{2})^{2}\}
\end{align*}%
\vspace{-1.5\baselineskip}%
\begin{equation}
\quad\quad\quad =\frac{1}{2}\mathbb{E}%
\{(X_{1}-X_{2})(X_{1}^{3}-X_{2}^{3})-3X_{3}(X_{1}-X_{2})(X_{1}^{2}-X_{2}^{2})+3X_{3}X_{4}(X_{1}-X_{2})^{2}\}.
\label{eq: Cov Fourth moment 4 rep}
\end{equation}

(\ref{eq: Cov Third moment 3 rep}) - (\ref{eq: Cov Fourth moment 4 rep})
still do not provide $D$-representations, because $X_{3}$ and $X_{4}$ are in
levels. This is done in the following propositions by using additional
replicates of $X$. A number of alternative expressions are provided in the
following proposition.

\begin{proposition}
\label{Th: D-representation of third central moment} %
\captionproposition{Proposition}{$D$-representation of third central moment}
Let $X$ be a random variable with mean $\mu _{X}$, variance $\sigma _{X}^{2}$%
, and finite third moment. If $X_{1}$, $X_{2}$, $X_{3}$ are three i.i.d.
replications of $X$, then%
\begin{eqnarray}
\mathbb{E}\{(X-\mu _{X})^{3}\} &=&\mathbb{E}\{(X_{1}-X_{3})(X_{1}-X_{2})^{2}%
\}=\frac{1}{2}\mathbb{E}\{(X_{1}-X_{2})[(X_{1}-X_{3})^{2}-(X_{2}-X_{3})^{2}]%
\}  \notag \\
&=&\frac{1}{2}\mathbb{E}\{(X_{1}-X_{2})^{2}[(X_{1}-X_{3})+(X_{2}-X_{3})]\}=%
\frac{1}{6}\mathbb{E}\{D_{1}^{(3)}D_{2}^{(3)}D_{3}^{(3)}\}  \notag \\
&=&\frac{9}{2}\mathbb{E}\{(X_{1}-\bar{X}^{(3)})(X_{2}-\bar{X}^{(3)})(X_{3}-%
\bar{X}^{(3)})\}\quad \quad  \label{eq: D-representation third moment}
\end{eqnarray}%
where\ \ $\bar{X}^{(3)}:=\,\underset{j=1}{\overset{3}{\sum }}X_{j}/3\;$and 
\begin{equation}
D_{i}^{(3)}:=\,\underset{j\neq i}{\sum }(X_{i}-X_{j})=3(X_{i}-\bar{X}%
^{(3)})\,,\quad i=1,\,2,\,3.  \label{eq: Sum of differences}
\end{equation}
\end{proposition}

As the variables $D_{i}^{(3)}$ are sums of pairwise differences, (\ref{eq:
D-representation third moment}) provides simple $D$-representations of the
third central moment of $X$. Note also that $\mathbb{E}\{(X-\mu _{X})^{3}\}$
can be rewritten as a function of deviations with respect to a
\textquotedblleft reference\textquotedblright\ value $X_{3}$:%
\begin{eqnarray}
\mathbb{E}\{(X-\mu _{X})^{3}\} &=&\frac{1}{2}\mathbb{E}%
\{[(X_{1}-X_{3})-(X_{2}-X_{3})\}[(X_{1}-X_{3})^{2}-(X_{2}-X_{3})^{2}]\} 
\notag \\
&=&\frac{1}{2}\mathbb{E}%
\{[(X_{1}-X_{3})-(X_{2}-X_{3})]^{2}[(X_{1}-X_{3})+(X_{2}-X_{3})]\}\,.
\end{eqnarray}%
We now consider the fourth central moment.

\begin{proposition}
\label{Th: D-representation of fourth central moment} 
\captionproposition{Proposition}{$D$-representation of fourth central
moment} Let $X$ be a random variable with mean $\mu _{X}$, variance $\sigma
_{X}^{2}$, and finite fourth moment. If $X_{1}$, $X_{2}$, $X_{3}$, $X_{4}$
four i.i.d. replications of $X$, then 
\begin{eqnarray}
\mathbb{E}\{(X-\mu _{X})^{4}\} &=&\frac{1}{2}\mathbb{E}\{(X_{1}-X_{2})^{4}%
\}-3\sigma _{X}^{4}=\frac{1}{2}\mathbb{E}\{(X_{1}-X_{2})^{4}\}-\frac{3}{4}(%
\mathbb{E}\{(X_{1}-X_{2})^{2}\})^{2}  \notag \\
&=&\mathbb{E}\big\{\frac{1}{2}(X_{1}-X_{2})^{4}-\frac{3}{4}%
(X_{1}-X_{2})^{2}(X_{3}-X_{4})^{2}\big\}\,.
\label{eq: D-representation fourth moment}
\end{eqnarray}
\end{proposition}

The last two identities in (\ref{eq: D-representation fourth moment})
provide simple $D$-representations of the fourth central moment of $X$. We
now apply the above results to skewness, kurtosis and excess kurtosis
coefficients: 
\begin{equation}
\mathrm{Sk}(X):=\frac{\mathbb{E}\{(X-\mu _{X})^{3}\}}{\sigma _{X}^{3}}\,,\;\;%
\mathrm{Kur}(X):=\frac{\mathbb{E}\{(X-\mu _{X})^{4}\}}{\sigma _{X}^{4}}%
\,,\;\;\mathrm{EKur}(X)=\mathrm{Kur}(X)-3\,.
\label{eq: Def Skewness Kurtosis}
\end{equation}%
$\mathrm{Kur}(X)$ is also called the \textquotedblleft Pearson
kurtosis\textquotedblright\ coefficient, while the excess kurtosis $\mathrm{%
EKur}(X)$ is the \textquotedblleft Fisher kurtosis\textquotedblright .

\begin{proposition}
\label{Th: D-representation of skewness and kurtosis} %
\captionproposition{Proposition}{$D$-representation of skewness and kurtosis}
Let $X$ be a random variable with mean $\mu _{X}$ and variance $\sigma
_{X}^{2}$, and $X_{1}$, $X_{2}$, $X_{3}$, $X_{4}$\ four i.i.d. replications
of $X$. If $X$ has finite third moment, then%
\begin{eqnarray}
\mathrm{Sk}(X) &=&\frac{\mathbb{E}\{(X_{1}-X_{2})(X_{1}^{2}-X_{2}^{2})\}}{%
2\sigma _{X}^{3}}-2\frac{\mu _{X}}{\sigma _{X}}=\frac{\mathbb{E}%
\{(X_{1}-X_{3})(X_{1}-X_{2})^{2}\}}{\sigma _{X}^{3}}  \notag \\
&=&\frac{\sqrt{8}\,\mathbb{E}\{(X_{1}-X_{3})(X_{1}-X_{2})^{2}\}}{(\mathbb{E}%
\{(X_{1}-X_{2})^{2}\})^{^{3/2}}}=\frac{\sqrt{2}}{3}\frac{\mathbb{E}%
\{D_{1}^{(3)}D_{2}^{(3)}D_{3}^{(3)}\}}{(\mathbb{E}\{(X_{1}-X_{2})^{2}%
\})^{^{3/2}}}
\end{eqnarray}%
where $D_{i}^{(3)}$ is defined by $(\ref{eq: Sum of differences})$. If $X$
has finite fourth moment, then%
\begin{equation}
\mathrm{Kur}(X)=\frac{2\mathbb{E}\{(X_{1}-X_{2})^{4}\}}{(\mathbb{E}%
\{(X_{1}-X_{2})^{2}\})^{2}}-3\,.
\end{equation}
\end{proposition}

If $X_{1}$ and $X_{2}$ are i.i.d. normal, it is easy to see that 
\begin{equation}
\frac{\mathbb{E}\{(X_{1}-X_{2})^{4}\}}{(\mathbb{E}\{(X_{1}-X_{2})^{2}\})^{2}}%
=\frac{12\sigma _{X}^{4}}{4\sigma _{X}^{4}}=3
\end{equation}%
so that $\mathrm{EKur}(X)=0$.

\begin{Tentative}
\end{Tentative}

\section{Higher-order Moments \label{Sec: Higher-order moments}}

\resetcountersSection

In this section, we study how higher-order central moments can be
represented in terms of pairwise differences between i.i.d. realizations of
a random variable. Consider a random variable $X$ with mean $\mu _{X}$ and
finite moments up to order $n\geq 1$. We denote the $k$-th central moment of 
$X$ by 
\begin{equation}
\mu _{k}:=\mathbb{E}\{(X-\mu _{X})^{k}\},\quad k=1,\ldots ,\,n\,.
\end{equation}%
Suppose that $X_{0},$ $X_{1},\ldots ,\,X_{n}$\ are i.i.d. replications of $X$%
, and define 
\begin{equation}
P_{k}:=\underset{i=1}{\overset{k}{\prod }}(X_{0}-X_{i})\,,\quad \tilde{X}%
_{i}:=X_{i}-\mu _{X}\,,\quad i=1,\ldots ,\,n\,.  \label{eq: P_j}
\end{equation}%
Since $\mathbb{E}\{\tilde{X}_{i}\}=\mu _{1}=0$, for $i=0,\ldots ,\,n$, the
independence of $\tilde{X}_{0},\ldots ,\,\tilde{X}_{n}$ entails:%
\begin{equation}
\mathbb{E}\{P_{k}\}=\mathbb{E}\big\{\underset{i=1}{\overset{k}{\prod }}(%
\tilde{X}_{0}-\tilde{X}_{i})\big\}=\mathbb{E}\{\tilde{X}_{0}^{k}\}=\mu
_{k}\,,\quad k=1,\ldots ,\,n\,.  \label{eq: E(P_j)}
\end{equation}%
In other words, $P_{k}$ can be viewed as an unbiased estimator of $\mu _{k}$%
, for $k=1,\ldots ,\,n$ . This immediately shows that all the central
moments of $X$ have a $D$-representation for moments up to order $n$,
provided $X$ has moments up to order $n$.

The function $P_{n}$ requires $n+1$ replications of $X$ to represent the $n$%
-th central moment of $X$. However, for $n=2$, $3,$ $4$, we have given
representations which only require $n$ replications (see Section \ref{Sec:
Skewness and kurtosis}). We will now show this is indeed feasible for $n\geq
5$. We first state some useful recursions.

\begin{proposition}
\label{Th: Recursions for higher-order moments} %
\captionproposition{Proposition}{Recursions for higher-order moments} Let $X$
be a random variable with mean $\mu _{X}$, and finite central moments $\mu
_{j}$ for $j=1,\ldots ,\,n+1$, where $n\geq 1.$ Then 
\begin{equation}
\mu _{n+1}=\mathsf{C}[X,\,(X-\mu _{X})^{n}]=\mathbb{E}\{X(X-\mu
_{X})^{n}\}-\mu _{X}\mu _{n}\,.  \label{eq: mu(n+1) cov}
\end{equation}%
If $X_{1}$, $X_{2}$, $X_{3}\;$are i.i.d. replications of $X$, then 
\begin{eqnarray}
\mu _{n+1} &=&\mathbb{E}\{(X_{1}-X_{2})(X_{1}-\mu _{X})^{n}\}  \notag \\
&=&\mathbb{E}\{(X_{1}-X_{3})(X_{1}-X_{2})^{n}\}-%
\sum_{j=2}^{n-1}(-1)^{j}C_{n}^{j}\,\mu _{j}\mu _{n+1-j}
\label{eq: recusion mu_n}
\end{eqnarray}%
where $C_{n}^{j}=n!/[j!(n-j)!]$ and, for $n+1$ even,%
\begin{equation}
\mu _{n+1}=\frac{1}{2}\big[\mathbb{E}\{(X_{1}-X_{2})^{n+1}\}-%
\sum_{j=2}^{n-1}(-1)^{j}C_{n+1}^{j}\,\mu _{j}\mu _{n+1-j}\big]\,.
\label{eq: recusion mu_n n even}
\end{equation}
\end{proposition}

When the summations in (\ref{eq: recusion mu_n})\thinspace -\thinspace (\ref%
{eq: recusion mu_n n even}) are empty (for $n=1$ and $n=2$), the
corresponding formulae yield [as expected from (\ref{eq: Var 3 rep}) and (%
\ref{eq: D-representation third moment})]:%
\begin{equation}
\mu _{2}=\mathbb{E}\{(X_{1}-X_{3})(X_{1}-X_{2})\}=\frac{1}{2}\mathbb{E}%
\{(X_{1}-X_{2})^{2}\}\,,\quad \mu _{3}=\mathbb{E}%
\{(X_{1}-X_{3})(X_{1}-X_{2})^{2}\}\,.
\end{equation}

Now, let $X_{1},\ldots ,\,X_{n}$ be i.i.d. replications of $X$ with $n\geq 5$
, and consider the following sequence of functions:%
\begin{equation}
\bar{\mu}_{1}=0\,,\quad \bar{\mu}_{2}(x_{1},\,x_{2})=\frac{1}{2}%
(x_{1}-x_{2})^{2}\,,  \label{eq: muhat(2)}
\end{equation}%
\begin{equation}
\bar{\mu}_{3}(x_{1},\,x_{2},\,x_{3})=(x_{1}-x_{3})(x_{1}-x_{2})^{2}\,,
\label{eq: muhat(3)}
\end{equation}%
\begin{equation}
\bar{\mu}_{4}(x_{1},\,x_{2},\,x_{3},\,x_{4})=\frac{1}{2}(x_{1}-x_{2})^{4}-%
\frac{3}{4}(x_{1}-x_{2})^{2}(x_{3}-x_{4})^{2},  \label{eq: mubar(4)}
\end{equation}%
and for $k\geq 4$,%
\begin{equation}
\bar{\mu}_{k+1}(x_{1},\ldots
,\,x_{k+1})=(x_{1}-x_{3})(x_{1}-x_{2})^{k}-\sum_{j=2}^{k-1}(-1)^{j}C_{k}^{j}%
\bar{\mu}_{j}(x_{1},\ldots ,\,x_{j})\bar{\mu}_{k+1-j}(x_{j+1},\ldots
,\,x_{k+1})\,.  \label{eq: mubar(k+1)}
\end{equation}%
If $k+1$ is even (with $k\geq 1$), we can also consider the functions%
\begin{eqnarray}
\tilde{\mu}_{k+1}(x_{1},\ldots ,\,x_{k+1}) &=&\frac{1}{2}\big[%
(x_{1}-x_{2})^{k+1}  \notag \\
&&-\sum_{j=2}^{k-1}(-1)^{j}C_{k+1}^{j}\bar{\mu}_{j}(x_{1},\ldots ,\,x_{j})%
\bar{\mu}_{k+1-j}(x_{j+1},\ldots ,\,x_{k+1})\big]\,.\quad \quad
\label{eq: mutilde(k)}
\end{eqnarray}%
From the above definitions, it is clear that the functions $\bar{\mu}%
_{k}(x_{1},\ldots ,\,x_{k})$ and $\tilde{\mu}_{k}(x_{1},\ldots ,\,x_{k})$
depend on their arguments only through differences $x_{i}-x_{j}$ ($%
i,j=1,\ldots ,\,k$). Note also that $\bar{\mu}_{j}$ can be replaced by $%
\tilde{\mu}_{j}$ in the summations of (\ref{eq: mubar(k+1)})\thinspace
-\thinspace (\ref{eq: mutilde(k)}) whenever $j$ is even.

\begin{proposition}
\label{Th: Recursive D-representations for higher-order moments} 
\captionproposition{Proposition}{Recursive $D$-representations for
higher-order moments} Let $X$ be a random variable with mean $\mu _{X}$, and
finite moments up to order $n+1$, where $n\geq 1$, and let $\bar{\mu}%
_{k}(x_{1},\ldots ,\,x_{k})$ and $\tilde{\mu}_{k}(x_{1},\ldots ,\,x_{k})$ be
defined by $(\ref{eq: muhat(2)})\,$-\thinspace $\,(\ref{eq: mutilde(k)}),$
for $k\geq 1.$ If $X_{1},\ldots ,\,X_{n+1}$ are i.i.d. replications of $X$,
then%
\begin{equation}
\mathbb{E}\{\bar{\mu}_{k+1}(X_{1},\ldots ,\,X_{k+1})\}=\mu _{k+1}\,,\quad 
\text{for }k=1,\ldots ,\,n\,,  \label{eq: E(mubar_k)}
\end{equation}%
and, if $k+1$ is even, 
\begin{equation}
\mathbb{E}\{\tilde{\mu}_{k+1}(X_{1},\ldots ,\,X_{k+1})\}=\mu _{k+1}\,,\quad 
\text{for\ }1\leq k\leq n\,.  \label{eq: E(mutilde_k)}
\end{equation}
\end{proposition}

The above proposition shows that the $k$-th central moment of $X$ has a $D$%
-representation which only requires $k$ replications of $X$ (provided $\mu
_{k}$ is finite).

\section{Conclusion \label{Sec: Conclusion}}

\resetcountersSection

In this paper, following an approach (apparently) initiated by \cite%
{Gini(1912)} for the variance, we have studied the general problem of
representing the central moments of a random variable in terms of pairwise
differences between i.i.d. replications, without reference to a location
parameter. We first gave pairwise-difference representations (defined
formally as $D$\emph{-representations}) for the third and fourth central
moments of a general random variable. These yield intuitive interpretations
of the familiar skewness and kurtosis coefficients. For general moments, we
have observed that central moments can be interpreted as covariances between
a random variable and powers of the same variable (a \emph{nonlinear}
relation). These then provide \emph{recursions} which link the $D$%
-representation of any moment to lower order ones, hence $D$-representations
for all central moments (as long as they are finite). Finally, it is of
interest to note that the formulas given in this paper can be interpreted as
applications of the \textquotedblleft coupling\textquotedblright\ method
[see \cite{Thorisson(2000)}], which opens the way to additional
applications. 

The above results show that unbiased estimators of central moments (with
unknown mean) can be obtained with a small number of observations: three
observations are sufficient for the third moment, for the fourth moment,
etc. Clearly, these results can be applied for the unbiased estimation of
central moments as well as testing moments. This raises a wide range of
associated statistical and comparative issues, which go beyond the scope of
this paper. Such applications will be considered in a follow-up paper.



 

\bibliographystyle{agsm}
\bibliography{refer1}

\newpage \appendix{}

\section{Appendix: Proofs \label{Proof of Propositions}}

\resetcountersSection

\begin{proofflexc}
\captionproofflexc{\propositionname}{Th: Covariance representation of third
and fourth central moments} The third central moment can be written as
follows: 
\begin{eqnarray}
\mathbb{E}\{(X-\mu _{X})^{3}\} &=&\mathbb{E}\{(X-\mu _{X})(X-\mu _{X})^{2}\}=%
\mathsf{C}[X,\,(X-\mu _{X})^{2}]  \notag \\
&=&\mathsf{C}[X,\,(X^{2}-2\mu _{X}X+\mu _{X}^{2})]=\mathsf{C}%
[X,\,(X^{2}-2\mu _{X}X)]\,.  \label{eq: E[(X-mu)^3]}
\end{eqnarray}%
Using (\ref{eq: Cov}) and (\ref{eq: Var}), we can write:%
\begin{eqnarray}
\mathsf{C}[X,\,(X-\mu _{X})^{2}] &=&\frac{1}{2}\mathbb{E}%
\{(X_{1}-X_{2})[(X_{1}-\mu _{X})^{2}-(X_{2}-\mu _{X})^{2}]\}  \notag \\
&=&\frac{1}{2}\mathbb{E}\{(X_{1}-X_{2})[(X_{1}^{2}-2\mu
_{X}X_{1})-(X_{2}^{2}-2\mu _{X}X_{2})]\}  \notag \\
&=&\frac{1}{2}\mathbb{E}\{(X_{1}-X_{2})[(X_{1}^{2}-X_{2}^{2})-2\mu
_{X}(X_{1}-X_{2})]\}  \notag \\
&=&\frac{1}{2}\mathbb{E}\{(X_{1}-X_{2})(X_{1}^{2}-X_{2}^{2})\}-\mu _{X}%
\mathbb{E}\{(X_{1}-X_{2})^{2}\}  \notag \\
&=&\frac{1}{2}\mathbb{E}\{(X_{1}-X_{2})(X_{1}^{2}-X_{2}^{2})\}-2\mu
_{X}\sigma _{X}^{2}\,.  \label{eq: Cov[X,(X-mu)^2]}
\end{eqnarray}%
(\ref{eq: Cov Third moment}) follows on combining (\ref{eq: E[(X-mu)^3]})
and (\ref{eq: Cov[X,(X-mu)^2]}). To show (\ref{eq: Cov Fourth moment}), we
proceed similarly. We have:%
\begin{eqnarray}
\mathbb{E}\{(X-\mu _{X})^{4}\} &=&\mathbb{E}\{(X-\mu _{X})(X-\mu _{X})^{3}\}=%
\mathsf{C}[X,\,(X-\mu _{X})^{3}]  \notag \\
&=&\mathsf{C}[X,\,(X^{3}-3X^{2}\mu _{X}+3X\mu _{X}^{2}-\mu _{X}^{3})]  \notag
\\
&=&\mathsf{C}[X,\,(X^{3}-3X^{2}\mu _{X}+3X\mu _{X}^{2})]\quad \quad
\label{eq: E[(X-mu)^4]}
\end{eqnarray}%
hence, by (\ref{eq: Cov}),%
\begin{eqnarray}
&&\mathsf{C}[X,\,(X-\mu _{X})^{3}]=\frac{1}{2}\mathbb{E}%
\{(X_{1}-X_{2})[(X_{1}-\mu _{X})^{3}-(X_{2}-\mu _{X})^{3}]\}  \notag \\
&=&\frac{1}{2}\mathbb{E}\{(X_{1}-X_{2})[(X_{1}^{3}-3X_{1}^{2}\mu
_{X}+3X_{1}\mu _{X}^{2}-\mu _{X}^{3})-\,(X_{2}^{3}-3X_{2}^{2}\mu
_{X}+3X_{2}\mu _{X}^{2}-\mu _{X}^{3})]\}  \notag \\
&=&\frac{1}{2}\mathbb{E}\{(X_{1}-X_{2})[(X_{1}^{3}-X_{2}^{3})-3\mu
_{X}(X_{1}^{2}-X_{2}^{2})+3\mu _{X}^{2}(X_{1}-X_{2})]\}  \notag \\
&=&\frac{1}{2}\mathbb{E}\{(X_{1}-X_{2})(X_{1}^{3}-X_{2}^{3})\}-\frac{3}{2}%
\mu _{X}\mathbb{E}\{(X_{1}-X_{2})(X_{1}^{2}-X_{2}^{2})\}+\frac{3}{2}\mu
_{X}^{2}\mathbb{E}\{(X_{1}-X_{2})^{2}\}  \notag \\
&=&\frac{1}{2}\mathbb{E}\{(X_{1}-X_{2})(X_{1}^{3}-X_{2}^{3})\}-3\mu _{X}\big[%
\mathbb{E}\{(X-\mu _{X})^{3}\}+\mu _{X}\mathbb{E}\{(X_{1}-X_{2})^{2}\}\big] 
\notag \\
&&+\frac{3}{2}\mu _{X}^{2}\mathbb{E}\{(X_{1}-X_{2})^{2}\}  \notag \\
&=&\frac{1}{2}\mathbb{E}\{(X_{1}-X_{2})(X_{1}^{3}-X_{2}^{3})\}-3\mu _{X}%
\mathbb{E}\{(X-\mu _{X})^{3}\}-\frac{3}{2}\mu _{X}^{2}\mathbb{E}%
\{(X_{1}-X_{2})^{2}\}  \notag \\
&=&\frac{1}{2}\mathbb{E}\{(X_{1}-X_{2})(X_{1}^{3}-X_{2}^{3})\}-3\mu _{X}%
\mathbb{E}\{(X-\mu _{X})^{3}\}-3\mu _{X}^{2}\sigma _{X}^{2}\,.
\label{eq: Cov[X,(X-mu)^3]}
\end{eqnarray}%
(\ref{eq: Cov Fourth moment}) follows on combining (\ref{eq: E[(X-mu)^4]})
and (\ref{eq: Cov[X,(X-mu)^3]}).
\end{proofflexc}

\quad

\begin{proofflexc}
\captionproofflexc{\propositionname}{Th: D-representation of third central
moment} Since $X_{1}$, $X_{2}$, $X_{3}$ are i.i.d. replications of $X$, we
have:%
\begin{equation*}
\mathbb{E}\{(X_{1}-X_{2})[(X_{1}-X_{3})^{2}-(X_{2}-X_{3})^{2}]\}=\mathbb{E}%
\{(X_{1}-X_{2})[(X_{1}^{2}-X_{2}^{2})-2X_{3}(X_{1}-X_{2})]\}
\end{equation*}%
$\quad $\vspace{-2.5\baselineskip}%
\begin{eqnarray}
&=&\mathbb{E}\{(X_{1}-X_{2})(X_{1}^{2}-X_{2}^{2})\}-2\mathbb{E}\{X_{3}\}%
\mathbb{E}\{(X_{1}-X_{2})^{2}\}  \notag \\
&=&\mathbb{E}\{(X_{1}-X_{2})(X_{1}^{2}-X_{2}^{2})\}-2\mu _{X}\mathbb{E}%
\{(X_{1}-X_{2})^{2}\}
\end{eqnarray}%
and, using (\ref{eq: Cov Third moment}), 
\begin{eqnarray}
\mathbb{E}\{(X-\mu _{X})^{3}\} &=&\frac{1}{2}\mathbb{E}%
\{(X_{1}-X_{2})(X_{1}^{2}-X_{2}^{2})\}-\mu _{X}\mathbb{E}\{(X_{1}-X_{2})^{2}%
\}  \notag \\
&=&\frac{1}{2}\mathbb{E}\{(X_{1}-X_{2})[(X_{1}-X_{3})^{2}-(X_{2}-X_{3})^{2}]%
\}  \notag \\
&=&\frac{1}{2}\{\mathbb{E}%
\{(X_{1}-X_{2})(X_{1}^{2}-X_{2}^{2})-2X_{3}(X_{1}-X_{2})^{2}]\}  \notag \\
&=&\frac{1}{2}\mathbb{E}\{(X_{1}-X_{2})^{2}[(X_{1}+X_{2})-2X_{3}]\}  \notag
\\
&=&\frac{1}{2}\mathbb{E}\{(X_{1}-X_{2})^{2}[(X_{1}-X_{3})+(X_{2}-X_{3})]\}\,.
\end{eqnarray}%
Now, set $\tilde{X}_{i}=X_{i}-\mu _{X}$, $i=1,$\thinspace $2$, $3$, so that $%
\mathbb{E}\{\tilde{X}_{i}\}=0$ for all $i$. Using the mutual independence of 
$\tilde{X}_{1}$, $\tilde{X}_{2}$, $\tilde{X}_{3}$, we can then write: 
\begin{eqnarray}
\mathbb{E}\{(X_{1}-X_{3})(X_{1}-X_{2})^{2}\} &=&\mathbb{E}\{(\tilde{X}_{1}-%
\tilde{X}_{3})(\tilde{X}_{1}-\tilde{X}_{2})^{2}\}=\mathbb{E}\{(\tilde{X}_{1}-%
\tilde{X}_{3})(\tilde{X}_{1}^{2}-2\tilde{X}_{1}\tilde{X}_{2}+\tilde{X}%
_{2}^{2})\}  \notag \\
&=&\mathbb{E}\{\tilde{X}_{1}^{3}\}=\mathbb{E}\{(X-\mu _{X})^{3}\}\,.\quad
\quad
\end{eqnarray}%
This establishes the first three identities of (\ref{eq: D-representation
third moment}). Further, 
\begin{equation}
D_{i}^{(3)}=\,\underset{j=1}{\overset{3}{\sum }}D_{ij}=\,\underset{j\neq i}{%
\sum }(\tilde{X}_{i}-\tilde{X}_{j})=2\tilde{X}_{i}-\underset{j\neq i}{\sum }%
\tilde{X}_{j}\,,\quad i=1,\,2,3,
\end{equation}%
where $\mathbb{E}\{\tilde{X}_{i}\tilde{X}_{j}X_{k}\}=0$ unless $i=j=k$,
hence 
\begin{gather}
\mathbb{E}\{D_{1}^{(3)}D_{2}^{(3)}D_{3}^{(3)}\}=\mathbb{E}\{(2\tilde{X}_{1}-%
\tilde{X}_{2}-\tilde{X}_{3})(2\tilde{X}_{2}-\tilde{X}_{1}-\tilde{X}_{3})(2%
\tilde{X}_{3}-\tilde{X}_{1}-\tilde{X}_{2})\}  \notag \\
=2\mathbb{E}\{\tilde{X}_{1}^{3}\}+2\mathbb{E}\{\tilde{X}_{2}^{3}\}+2\mathbb{E%
}\{\tilde{X}_{3}^{3}\}=6\mathbb{E}\{(X-\mu _{X})^{3}\}\,.
\end{gather}%
The last identity of (\ref{eq: D-representation third moment}) follows on
observing that 
\begin{equation}
X_{i}-\bar{X}^{(3)}=\frac{1}{3}(3X_{i}-\underset{j=1}{\overset{3}{\sum }}%
X_{j})=\frac{1}{3}\underset{j\neq i}{\sum }(X_{i}-X_{j})=\frac{1}{3}%
D_{i}^{(3)}.
\end{equation}
\end{proofflexc}

\quad

\begin{Tentative}
\end{Tentative}

\begin{proofflexc}
\captionproofflexc{\propositionname}{Th: D-representation of fourth central
moment} Set $\tilde{X}_{i}=X_{i}-\mu _{X}$, $i=1,\ldots ,\,4$, and consider
the binomial expansion:%
\begin{eqnarray}
(X_{1}-X_{2})^{4} &=&(\tilde{X}_{1}-\tilde{X}_{2})^{4}=%
\sum_{j=0}^{4}(-1)^{j}C_{4}^{j}\tilde{X}_{1}^{4-j}\tilde{X}_{2}^{j}  \notag
\\
&=&(\tilde{X}_{1}^{4}+\tilde{X}_{2}^{4})-4(\tilde{X}_{1}^{3}\tilde{X}_{2}+%
\tilde{X}_{1}\tilde{X}_{2}^{3})+6\tilde{X}_{1}^{2}\tilde{X}_{2}^{2}\,.
\end{eqnarray}%
Taking the expected value on both sides of the above equation, we get:%
\begin{eqnarray}
\mathbb{E}\{(X_{1}-X_{2})^{4}\} &=&2\mathbb{E}\{(X-\mu _{X})^{4}\}-4(\mathbb{%
E}\{\tilde{X}_{1}^{3}\}\mathbb{E}\{\tilde{X}_{2}\}+\mathbb{E}\{\tilde{X}%
_{1}\}\mathbb{E}\{\tilde{X}_{2}^{3}\})+6\mathbb{E}\{\tilde{X}_{1}^{2}\}%
\mathbb{E}\{\tilde{X}_{2}^{2}\}  \notag \\
&=&2\mathbb{E}\{(X-\mu _{X})^{4}\}+6\sigma _{X}^{4}=2\mathbb{E}([(X-\mu
_{X})^{4}\}+\frac{3}{2}(\mathbb{E}\{(X_{1}-X_{2})^{2}\})^{2}  \notag \\
&=&2\mathbb{E}\{(X-\mu _{X})^{4}\}+\frac{3}{2}\mathbb{E}\{(X_{1}-X_{2})^{2}\}%
\mathbb{E}\{(X_{3}-X_{4})^{2}\}  \notag \\
&=&2\mathbb{E}\{(X-\mu _{X})^{4}\}+\frac{3}{2}\mathbb{E}%
\{(X_{1}-X_{2})^{2}(X_{3}-X_{4})^{2}\}
\end{eqnarray}%
hence%
\begin{eqnarray}
\mathbb{E}\{(X-\mu _{X})^{4}\} &=&\frac{1}{2}\mathbb{E}\{(X_{1}-X_{2})^{4}%
\}-3\sigma _{X}^{4}  \notag \\
&=&\frac{1}{2}\mathbb{E}\{(X_{1}-X_{2})^{4}\}-\frac{3}{4}(\mathbb{E}%
\{(X_{1}-X_{2})^{2}\})^{2}  \notag \\
&=&\frac{1}{2}\mathbb{E}\{(X_{1}-X_{2})^{4}\}-\frac{3}{4}\mathbb{E}%
\{(X_{1}-X_{2})^{2}(X_{3}-X_{4})^{2}\}  \notag \\
&=&\frac{1}{2}\mathbb{E}\{(X_{1}-X_{2})^{4}-\frac{3}{2}%
(X_{1}-X_{2})^{2}(X_{3}-X_{4})^{2}\}\,.
\end{eqnarray}%
This completes the proof of (\ref{eq: D-representation fourth moment}).
\end{proofflexc}

\quad

\begin{proofflexc}
\captionproofflexc{\propositionname}{Th: D-representation of skewness and
kurtosis} If\ $X$ has finite third moment, we get from Propositions \ref{Th:
Covariance representation of third and fourth central moments} and \ref{Th:
D-representation of third central moment}:%
\begin{eqnarray}
\mathrm{Sk}(X) &=&\frac{\mathbb{E}\{(X-\mu _{X})^{3}\}}{\sigma _{X}^{3}}=%
\frac{(\mathbb{E}\{(X_{1}-X_{2})(X_{1}^{2}-X_{2}^{2})\}/2)-2\mu _{X}\sigma
_{X}^{2}}{\sigma _{X}^{3}}  \notag \\
&=&\frac{\mathbb{E}\{(X_{1}-X_{3})(X_{1}-X_{2})^{2}\}}{\sigma _{X}^{3}}=%
\frac{\mathbb{E}\{(X_{1}-X_{3})(X_{1}-X_{2})^{2}\}}{(\sigma _{X}^{2})^{3/2}}
\notag \\
&=&\frac{\mathbb{E}\{(X_{1}-X_{3})(X_{1}-X_{2})^{2}\}}{(\mathbb{E}%
\{(X_{1}-X_{2})^{2}\}/2)^{3/2}}=\frac{\sqrt{8}\mathbb{E}%
\{(X_{1}-X_{3})(X_{1}-X_{2})^{2}\}}{(\mathbb{E}\{(X_{1}-X_{2})^{2}\})^{3/2}}
\notag \\
&=&\frac{\sqrt{8}}{6}\frac{\mathbb{E}\{D_{1}^{(3)}D_{2}^{(3)}D_{3}^{(3)}\}}{(%
\mathbb{E}\{(X_{1}-X_{2})^{2}\})^{3/2}}=\frac{\sqrt{2}}{3}\frac{\mathbb{E}%
\{D_{1}^{(3)}D_{2}^{(3)}D_{3}^{(3)}\}}{(\mathbb{E}\{(X_{1}-X_{2})^{2}%
\})^{3/2}}\,.
\end{eqnarray}%
If $X$ has finite fourth moment, we get from Propositions \ref{Th:
Covariance representation of third and fourth central moments} and \ref{Th:
D-representation of fourth central moment}:%
\begin{eqnarray}
\mathrm{Kur}(X) &=&\frac{\mathbb{E}\{(X-\mu _{X})^{4}\}}{\sigma _{X}^{4}} 
\notag \\
&=&\frac{\mathbb{E}\{(X_{1}-X_{2})^{4}\}-6\sigma _{X}^{4}}{2\sigma _{X}^{4}}=%
\frac{\mathbb{E}\{(X_{1}-X_{2})^{4}\}}{2\sigma _{X}^{4}}-3  \notag \\
&=&\frac{\mathbb{E}\{(X_{1}-X_{2})^{4}\}}{2(\mathbb{E}\{(X_{1}-X_{2})^{2}%
\}/2)^{2}}-3=\frac{2\mathbb{E}\{(X_{1}-X_{2})^{4}\}}{(\mathbb{E}%
\{(X_{1}-X_{2})^{2}\})^{2}}-3\,.
\end{eqnarray}
\end{proofflexc}

\begin{proofflexc}
\captionproofflexc{\propositionname}{Th: Recursions for higher-order moments}
By definition, for $n\geq 1$,%
\begin{eqnarray}
\mu _{n+1} &=&\mathbb{E}\{(X-\mu _{X})(X-\mu _{X})^{n}\}=\mathsf{C}%
[X,\,(X-\mu _{X})^{n}]  \notag \\
&=&\mathbb{E}\{X(X-\mu _{X})^{n}\}-\mu _{X}\mu _{n}
\end{eqnarray}%
which yields (\ref{eq: mu(n+1) cov}). Let $X_{1}$, $X_{2}$, $X_{3}$ be
i.i.d. replications of $X$. Then, the above identity entails:%
\begin{eqnarray}
\mu _{n+1} &=&\mathbb{E}\{X_{1}(X_{1}-\mu _{X})^{n}\}-\mathbb{E}\{X_{2}\}%
\mathbb{E}\{(X_{1}-\mu _{X})^{n}\}  \notag \\
&=&\mathbb{E}\{X_{1}(X_{1}-\mu _{X})^{n}\}-\mathbb{E}\{X_{2}(X_{1}-\mu
_{X})^{n}\}  \notag \\
&=&\mathbb{E}\{(X_{1}-X_{2})(X_{1}-\mu _{X})^{n}\}\,.
\end{eqnarray}%
Setting $\tilde{X}_{j}:=X_{j}-\mu _{X}\,,\;j=1,$\thinspace $2\,,$ $3$, and
using the binomial theorem, we can write:%
\begin{eqnarray}
(X_{1}-X_{2})^{n} &=&(\tilde{X}_{1}-\tilde{X}_{2})^{n}=%
\sum_{j=0}^{n}C_{n}^{j}\,\tilde{X}_{1}^{n-j}(-\tilde{X}_{2})^{j}  \notag \\
&=&\sum_{j=0}^{n}(-1)^{j}C_{n}^{j}\,\tilde{X}_{1}^{n-j}\tilde{X}_{2}^{j}=%
\tilde{X}_{1}^{n}+(-1)^{n}\tilde{X}_{2}^{n}+%
\sum_{j=1}^{n-1}(-1)^{j}C_{n}^{j}\,\tilde{X}_{1}^{n-j}\tilde{X}_{2}^{n}\quad
\quad  \label{eq: (X_1-X_2)^n}
\end{eqnarray}%
and, multiplying the above equation by $(X_{1}-X_{3})$, 
\begin{gather}
(X_{1}-X_{3})(X_{1}-X_{2})^{n}=(\tilde{X}_{1}-\tilde{X}_{3})(\tilde{X}_{1}-%
\tilde{X}_{2})^{n}  \notag \\
=(\tilde{X}_{1}-\tilde{X}_{3})[\tilde{X}_{1}^{n}+(-1)^{n}\tilde{X}_{2}^{n}]+(%
\tilde{X}_{1}-\tilde{X}_{3})\sum_{j=1}^{n-1}(-1)^{j}C_{n}^{j}\,\tilde{X}%
_{1}^{n-j}\tilde{X}_{2}^{j}\,.  \label{eq: (X_1-X_3)(X_1-X_2)^n}
\end{gather}%
Taking the expected value on both sides of (\ref{eq: (X_1-X_3)(X_1-X_2)^n})
and using the fact that $\mathbb{E}\{\tilde{X}_{3}\}=0$ and $\mathbb{E}\{%
\tilde{X}_{1}^{k}\}=\mathbb{E}\{\tilde{X}_{2}^{k}\}=\mu _{k}$, we get:%
\begin{eqnarray}
\mathbb{E}\{(X_{1}-X_{3})(X_{1}-X_{2})^{n}\} &=&\mathbb{E}\{\tilde{X}%
_{1}^{n+1}\}+\sum_{j=1}^{n-1}(-1)^{j}C_{n}^{j}\,\mathbb{E}\{\tilde{X}%
_{1}^{n+1-j}\}\mathbb{E}\{\tilde{X}_{2}^{j}\}  \notag \\
&=&\mu _{n+1}+\sum_{j=1}^{n-1}(-1)^{j}C_{n}^{j}\,\mu _{n+1-j}\mu _{j}
\end{eqnarray}%
hence, noting that $\mu _{1}=0$, 
\begin{equation}
\mu _{n+1}=\mathbb{E}\{(X_{1}-X_{3})(X_{1}-X_{2})^{n}\}-%
\sum_{j=2}^{n-1}(-1)^{j}C_{n}^{j}\,\mu _{j}\mu _{n+1-j}\,.
\end{equation}%
This establishes (\ref{eq: recusion mu_n}). When $n$ is even, we can take
the expected value on both sides of (\ref{eq: (X_1-X_2)^n}):%
\begin{eqnarray}
\mathbb{E}\{(X_{1}-X_{2})^{n}\} &=&\mathbb{E}\{\tilde{X}_{1}^{n}+(-1)^{n}%
\tilde{X}_{2}^{n}\}+\sum_{j=1}^{n-1}(-1)^{j}C_{n}^{j}\,\mathbb{E}\{\tilde{X}%
_{1}^{n-j}\}\mathbb{E}\{\tilde{X}_{2}^{j}\}  \notag \\
&=&2\mu _{n}+\sum_{j=1}^{n-1}(-1)^{j}C_{n}^{j}\,\mu _{j}\mu _{n-j}=2\mu
_{n}+\sum_{j=2}^{n-2}(-1)^{j}C_{n}^{j}\,\mu _{j}\mu _{n-j}\quad \quad
\end{eqnarray}%
hence 
\begin{equation}
\mu _{n}=\frac{1}{2}\big[\mathbb{E}\{(X_{1}-X_{2})^{n}\}-%
\sum_{j=2}^{n-2}(-1)^{j}C_{n}^{j}\,\mu _{j}\mu _{n-j}\big]\,
\end{equation}%
which yields (\ref{eq: recusion mu_n n even}).
\end{proofflexc}

\quad

\begin{proofflexc}
\captionproofflexc{\propositionname}{Th: Recursive D-representations for
higher-order moments} To show (\ref{eq: E(mubar_k)}), we first observe that
the result holds for $k=1\,$, $2$, $3$ [by (\ref{eq: Var}) and Propositions %
\ref{Th: D-representation of third central moment} - \ref{Th:
D-representation of skewness and kurtosis}]. We can then proceed by
mathematical induction. Suppose that (\ref{eq: E(mubar_k)}) holds for values
of $k$ such that $1\leq k<n$, \emph{i.e.} 
\begin{equation}
\mathbb{E}\{\bar{\mu}_{k}(X_{1},\ldots ,\,X_{k})\}=\mu _{k}\,,\;\text{for }%
1\leq k<n\,.
\end{equation}%
For $k\geq 4$ , we replace $(x_{1},\ldots ,\,x_{k+1})$ by $(X_{1},\ldots
,\,X_{k+1})$ in (\ref{eq: mubar(k+1)}) and we take the expected value:%
\begin{eqnarray*}
\mathbb{E}\{\bar{\mu}_{k+1}(X_{1},\ldots ,\,X_{k+1})\} &=&\mathbb{E}%
\{(X_{1}-X_{3})(X_{1}-X_{2})^{k}\} \\
&&-\sum_{j=2}^{k-1}(-1)^{j}C_{k}^{j}\mathbb{E}\{\bar{\mu}_{j}(X_{1},\ldots
,\,X_{j})\bar{\mu}_{k+1-j}(X_{j+1},\ldots ,\,X_{k+1})\}
\end{eqnarray*}%
\quad \vspace{-2\baselineskip}%
\begin{eqnarray}
&=&\mathbb{E}\{(X_{1}-X_{3})(X_{1}-X_{2})^{k}\}  \notag \\
&&-\sum_{j=2}^{k-1}(-1)^{j}C_{k}^{j}\mathbb{E}\{\bar{\mu}_{j}(X_{1},\ldots
,\,X_{j})\}\mathbb{E}\{\bar{\mu}_{k+1-j}(X_{j+1},\ldots ,\,X_{k+1})\}  \notag
\\
&=&\mathbb{E}\{(X_{1}-X_{3})(X_{1}-X_{2})^{k}\}-%
\sum_{j=2}^{k-1}(-1)^{j}C_{k}^{j}\mu _{j}\mu _{k+1-j}
\end{eqnarray}%
where the last identity follows from the independence of $(X_{1},\ldots
,\,X_{j})$ and $(X_{j+1},\ldots ,\,X_{k+1})$. By (\ref{eq: recusion mu_n}),
this implies 
\begin{equation}
\mathbb{E}\{\bar{\mu}_{k+1}(X_{1},\ldots ,\,X_{k+1})\}=\mu _{k+1}\,.
\end{equation}%
This establishes (\ref{eq: E(mubar_k)}). If $k+1$ is even, we get from (\ref%
{eq: mutilde(k)}) and (\ref{eq: recusion mu_n n even}):%
\begin{eqnarray*}
\mathbb{E}\{\tilde{\mu}_{k+1}(X_{1},\ldots ,\,X_{k+1})\} &=&\frac{1}{2}\big[%
\mathbb{E}\{(X_{1}-X_{2})^{k+1}\} \\
&&-\sum_{j=2}^{k-1}(-1)^{j}C_{k+1}^{j}\mathbb{E}\{\bar{\mu}_{j}(X_{1},\ldots
,\,X_{j})\bar{\mu}_{k+1-j}(X_{j+1},\ldots ,\,X_{k+1})\}\big]
\end{eqnarray*}%
\begin{equation*}
=\frac{1}{2}\big[\mathbb{E}\{(X_{1}-X_{2})^{k+1}\}-%
\sum_{j=2}^{k-1}(-1)^{j}C_{k+1}^{j}\mathbb{E}\{\bar{\mu}_{j}(X_{1},\ldots
,\,X_{j})\}\mathbb{E}\{\bar{\mu}_{k+1-j}(X_{j+1},\ldots ,\,X_{k+1})\}\big]
\end{equation*}%
\begin{equation}
=\frac{1}{2}\big[\mathbb{E}\{(X_{1}-X_{2})^{k+1}\}-%
\sum_{j=2}^{k-1}(-1)^{j}C_{k+1}^{j}\mu _{j}\mu _{k+1-j}\big]=\mu _{k+1}\,.
\end{equation}%
This completes the proof of (\ref{eq: E(mutilde_k)}).
\end{proofflexc}

\begin{Tentative}
\end{Tentative}

\begin{Tentative}
\end{Tentative}

\end{document}